\pdfoutput=1

\documentclass[12pt, reqno]{article}
\usepackage{epsfig}
\usepackage{amssymb}
\usepackage{amsmath}
\usepackage{mathrsfs}
\usepackage{cite}
\usepackage{hyperref}

\newcommand{\et}{\zeta}

\newcommand{\be}{\begin{equation}}
\newcommand{\ee}{\end{equation}}

\renewcommand{\tilde}{\widetilde}
\newcommand{\smallminus}{-}
\newcommand{\smallplus}{+}

\makeindex
\begin{document}

\begin{titlepage}

\begin{flushright}
\vspace{-3cm}PUPT-2326\vspace{1cm}
\end{flushright}
\begin{center}

{\Large \bf Local Spacetime Physics from the Grassmannian}

\vspace{0.2cm}

{\bf N. Arkani-Hamed$^a$, J. Bourjaily$^{a,c}$, F. Cachazo$^{b,a}$, J. Trnka$^{a,c}$}

\vspace{.1cm}

{\it $^{a}$ School of Natural Sciences, Institute for Advanced Study, Princeton, NJ 08540, USA}

{\it $^{b}$ Perimeter Institute for Theoretical Physics, Waterloo, Ontario N2J W29, CA}

{\it $^{c}$ Department of Physics, Princeton University, Princeton, NJ 08544, USA}
\end{center}
\vskip.3in

\begin{abstract}
A duality has recently been conjectured between all leading singularities of $n$-particle N$^{k\smallminus2}$MHV scattering amplitudes in $\mathcal{N} = 4$ SYM and the residues of a contour integral with a natural measure over the Grassmannian $G(k,n)$. In this note we show that a simple contour deformation converts the sum of Grassmannian residues associated with the BCFW expansion of NMHV tree amplitudes to the CSW expansion of the same amplitude.
We propose that for general $k$ the same deformation yields the $(k-2)$ parameter Risager expansion. We establish this equivalence for all $\overline{{\rm MHV}}$ amplitudes and show that the Risager degrees of freedom are non-trivially determined by the GL$(k-2)$ ``gauge" degrees of freedom in the Grassmannian. The Risager expansion is known to recursively construct the CSW expansion for all tree amplitudes, and given that the CSW expansion follows directly from the (super) Yang-Mills Lagrangian in light-cone gauge, this contour deformation allows us to directly see the emergence of local space-time physics from the \mbox{Grassmannian}.
\end{abstract}

\vspace{\fill}

\end{titlepage}

\bigskip
\bigskip

\section{${\cal N} = 4$ SYM and the Grassmannian}

A dual formulation for the S-Matrix of ${\cal N} = 4$ SYM has recently been proposed \cite{ArkaniHamed:2009dn}, where the leading singularities of the $n$-particle N$^{k\smallminus2}$MHV amplitudes---to all orders in perturbation theory---are associated with a remarkably simple integral over the Grassmannian $G(k,n)$: \vspace{-0.2cm}
\be
\label{first}
{\cal L}_{n,k}({\cal W}) = \frac{1}{{\rm vol} [{\rm GL}(k)]} \int \frac{d^{\, k \times n} C_{\alpha a}}{(12 \cdots k) (23 \cdots k\smallplus1) \cdots (n1 \cdots k\smallminus1)} \prod_{\alpha=1}^k \delta^{4|4}(C_{\alpha a} {\cal W}_a).\vspace{-0.2cm}
\ee
Let us quickly review the notation appearing in (\ref{first}). First, the Grassmannian is the space of $k$-planes in $n$ dimensions, an element of which can be represented by a collection of $k$ $n$-vectors in the $n$-dimensional space whose span specifies the plane. These vectors can be put together into the $k \times n$ matrix $C_{\alpha a}$, where $\alpha = 1,\ldots, k$ and $a = 1,\ldots, n$. With this, we write
\be
(m_1 m_2 \cdots m_k) = \epsilon^{\alpha_1 \cdots \alpha_k} C_{\alpha_1 m_1} \cdots C_{\alpha_k m_k}
\ee
for the minor of the $k \times n$ matrix $C_{\alpha a}$ made from the columns $(m_1, \cdots, m_k)$.
Since any $k \times k$ linear transformation on these $k$ vectors leaves the $k$-plane invariant, there is a GL($k$) ``gauge symmetry" $C_{\alpha a} \mapsto L_{\alpha}^{\beta} C_{\beta a}$; our integral is ``gauge-fixed" by dividing by the volume of GL$(k)$. The amplitude is given in dual twistor space, ${\cal W}_a = (\tilde \mu_a, \tilde \lambda_a| \tilde \eta_a)$, where $\tilde \mu_a$ is the (half-Fourier transform) conjugate of $\tilde \lambda_a$, and $\tilde \eta_a$ is a SUSY Grassmann parameter.

This expression can be trivially transformed back to momentum space---the only dependence is in the $\delta^{4|4}(C_{\alpha a} {\cal W}_a)$ factor, which transforms into\vspace{-0.25cm}
\be
\label{twoplanes}
\delta^{4|4}(C_{\alpha a} {\cal W}_a) \to \int d^{2 \times k} \rho^\alpha \prod_{a=1}^n \delta^2(\rho^\alpha C_{\alpha a} - \lambda_a) \times \prod_{\alpha=1}^k \delta^2(C_{\alpha a} \tilde \lambda_a) \times \delta^4(C_{\alpha a} \tilde \eta_a).\vspace{-0.25cm}
\ee
In words, this equation embodies a simple new way of thinking about momentum conservation. The kinematical data is given by specifying $n$ individual $\lambda_a$'s and $\tilde \lambda_a$'s, each of which has two Lorentz indices. We can think of each (Lorentz) component as specifying some $n$-vector in the $n$-dimensional space of particle labels. Actually, given that the Lorentz group is SL$(2) \times$SL$(2)$, the Lorentz-invariant statement is that there is a two-plane $\lambda$ and another two-plane $\tilde \lambda$; momentum conservation $\sum \lambda_a \tilde \lambda_a = 0$ is the statement that the two-planes $\lambda$ and $\tilde \lambda$ are orthogonal. \mbox{Equation (\ref{twoplanes})} interprets this in a different way, by introducing an auxiliary object---the $k$-plane $C$---and forcing $C$ to contain the $\lambda$-plane (the first factor) and be orthogonal to the $\tilde \lambda$-plane (the second factor).
\begin{figure}[b]\label{fig1}\vspace{-1.9cm}\centering\hspace{1.cm}\includegraphics[scale=1]{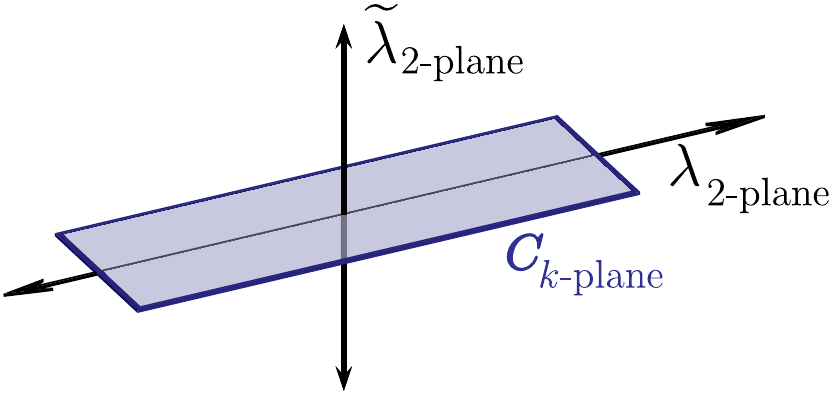}\vspace{-1.05cm}\end{figure}
\newpage

The final, Grassmann $\delta$-function in equation (\ref{twoplanes}) ensures that the object is invariant under all of GL$(k)$ (and not just SL$(k)$). In fact, we could have motivated the entire construction leading to \mbox{equation (\ref{first})} from this picture of momentum conservation: the measure in the integral over the Grassmannian is simply the nicest GL$(k)$-invariant one with manifest cyclic symmetry. Note also that while (\ref{first}) makes superconformal invariance manifest, the momentum-space form involving (\ref{twoplanes}) makes parity manifest: the action of parity is just the obvious map between $G(k,n)$ and $G(n\smallminus k,n)$. This can be seen explicitly by choosing a natural gauge-fixing of GL$(k)$, where $k$ of the columns of $C$ are set to an orthonormal basis, corresponding to the ``link-representation" \cite{ArkaniHamed:2009dn,ArkaniHamed:2009si}.

The geometric picture of momentum conservation motivates yet another representation of ${\cal L}_{n,k}$, which makes {\it dual} superconformal invariance manifest \cite{ArkaniHamed:2009vw,Mason:2009qx}.
Since momentum conservation requires that the $C$-plane contains the $\lambda$ two-plane, it is possible to re-write the integral as one over only the space of $(k-2)$-planes, $D$, which are complementary to $\lambda$ in $C$. This can be done using a gauge-fixing of GL$(k)$ which forces the first two rows of the $C$-matrix to coincide with the $\lambda$-plane---thereby manifestly encoding the fact that the Grassmannian includes the $\lambda$-plane. A further linear transformation maps $k \times k$ minors to $(k-2) \times (k-2)$ minors, and we find that we can write
\be
{\cal L}_{n,k}(\lambda,\tilde \lambda, \tilde \eta) = \frac{\delta^4(\sum_a \lambda_a \tilde \lambda_a) \delta^8(\sum_a \lambda_a \tilde \eta_a)}{\langle 1 2 \rangle \langle 2 3 \rangle \cdots \langle n 1 \rangle} \times {\cal R}_{n,k},
\ee
where
\be
\label{momtwist}{\cal R}_{n,k}({\cal Z}) = \frac{1}{{\rm vol} [{\rm GL}(k\smallminus2)]} \int \frac{d^{\, (k\smallminus2)\times n} D_{\hat{\alpha} a}}{(12\,\cdots\,\,k\smallminus2)(23\,\cdots\,\,k\smallminus 1) \cdots (n1\,\cdots\,\,k\smallminus3)} \prod_{\hat{\alpha}=1}^{k-2}\delta^{4|4}(D_{\hat{\alpha} a} {\cal Z}_a).
\ee
Here, the $\mathcal{Z}_a$ are the ``momentum-twistor" variables introduced by Hodges \cite{Hodges:2009hk}, which are the most natural variables with which to discuss {\it dual} superconformal invariance. External particles are associated with points $x_a$ in the dual space, with $p_a = x_{a+1} - x_a$. The point $x_a$ is associated with a line in {\it its} associated momentum-``twistor space"; and since $x_a - x_{a+1}$ is null, the line in momentum-twistor space associated with $x_a$ intersects the line associated with $x_{a+1}$. Therefore, we can associate $x_a$ with a canonical pair of momentum-twistors $(\mathcal{Z}_a,\mathcal{Z}_{a-1})$ defined by the intersection of lines. This is illustrated in the figure below.
\begin{figure}[t]
\vspace{-0.4cm}\centering\includegraphics[scale=1]{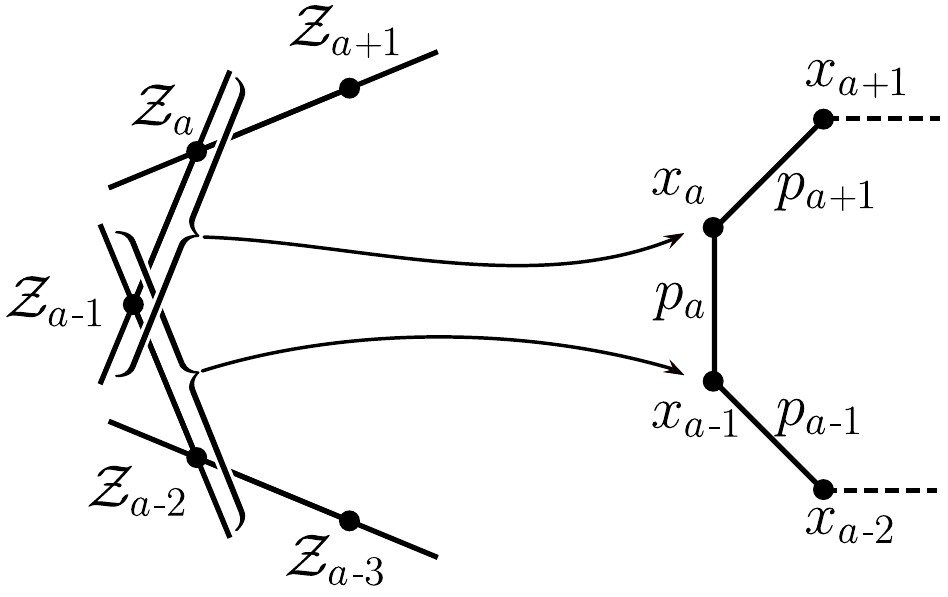}\vspace{-0.5cm}
\end{figure}
The momentum twistor $\mathcal{Z}_a$ is composed of ${\cal Z}_a = (\mu_a, \lambda_a|\eta_a)$, where the variables $\tilde \lambda_a, \tilde \eta_a$ are determined by $\mu_a, \eta_a$. Explicitly, they are given by
\be
\begin{split}
\tilde{\lambda}_a =\,& \frac{\langle a-1\,\, a\rangle \mu_{a+1} +
\langle a\,\, a+1\rangle \mu_{a-1}+\langle a+1\,\, a-1\rangle
\mu_a}{\langle a-1\,\, a\rangle\langle a\,\, a+1\rangle} \\
\tilde \eta_a  =\,& \frac{\langle a-1\,\, a\rangle \eta_{a+1} +
\langle a\,\, a+1\rangle \eta_{a-1}+\langle a+1\,\, a-1\rangle
\eta_a}{\langle a-1\,\, a\rangle\langle a\,\, a+1\rangle}
\end{split}.
\ee

Dual superconformal transformations \cite{Mason:2009qx,Hodges:2009hk,Brandhuber:2008pf,Drummond:2008vq} are just linear transformations of the ${\cal Z}_a$, which is a manifest symmetry of \mbox{equation (\ref{momtwist})}, just as ordinary superconformal transformations are linear transformations on ${\cal W}_a$ making them a manifest symmetry of \mbox{equation (\ref{first})}. Thus, \mbox{equation (\ref{first})} makes {\it all} the important symmetries of ${\cal N}=4$ SYM amplitudes manifest.

The momentum-space formula for ${\cal L}_{n,k}$ is to be interpreted as a contour integral in \mbox{$(k-2)\times (n-k-2)$} variables, which can be thought of as specifying the unfixed degrees of freedom of a $(k-2)$-plane orthogonal to both the $\tilde \lambda$- and $\lambda$-planes. In \cite{ArkaniHamed:2009dn}, evidence was given that the residues of the integrand are associated with leading singularities up to 2 loops, motivating the conjecture that {\it all} leading singularities are contained as residues. This conjecture carries even more weight given the realization that all the residues are both superconformal {\it and} dual superconformal invariant, which further means they are invariant under the full Yangian symmetry \cite{Drummond:2008vq}. Leading singularities are data associated with scattering amplitudes that are free of IR divergences---at loop level, they can be thought of as being associated with loop integrals over compact contours---and should therefore reflect all the symmetries of the theory. In fact, the residues of our object can be thought of as generating (likely all) Yangian invariants that are algebraic functions of the external spinor-helicity variables. Furthermore, as emphasized in \cite{ArkaniHamed:2009dn}, higher-dimensional residue theorems encode highly non-trivial relations between these invariants, many of which have striking physical interpretations such as loop-level infrared equations.

It is clear that there is an enormous amount of fascinating structure to be uncovered in the properties of the individual residues of ${\cal L}_{n,k}$, since they are invariants of the most remarkable integrable structure we have ever seen in physics! Recent work \cite{Kaplan:2009mh,Bullimore:2009cb} as well as work to appear \cite{ABCCK:2010} gives strong evidence that infinite classes of all-loop leading singularities are indeed contained amongst the residues of ${\cal L}_{n,k}$.

There is however something even more remarkable than the properties of residues taken individually: they can be combined in such a way as to produce amplitudes with a local space-time interpretation. Consider for instance NMHV tree amplitudes ($k=3$). A given residue is associated with putting $(k-2)(n-k-2) = (n-5)$ minors to zero, which can be labeled as $(m_1) \cdots (m_{n-5})$, where $(m)$ denotes that the minor $(m\,\,\,m\smallplus1\,\,\,m\smallplus2)$ has been set to zero. In \cite{ArkaniHamed:2009dn}, it was shown that a natural BCFW expansion for the NMHV amplitudes is given by a sum of residues
\be
\label{nmhvcont}
M^{{\rm BCFW}}_{n,{\rm NMHV}} =\Large\sum\normalsize\underbrace{(o_1)(e_2)(o_3)\,\,\cdots\phantom{+}}_{\text{\normalsize{$n-5$ terms}}}
\ee
where the sum is over all strictly-increasing series of $(n-5)$ alternating odd ($o$) and even ($e$) integers; to be explicit the 6-,7- and 8-particle amplitudes are given by
\begin{equation}
\begin{split}
M^{{\rm BCFW}}_{6,{\rm NMHV}}\,\,&=\phantom{+}(1) + (3) + (5);\\
M^{{\rm BCFW}}_{7,{\rm NMHV}}\,\,&=\phantom{+}(1)(2) + (1)(4) + (1)(6) + (3)(4) + (3)(6) + (5)(6);\\
M^{{\rm BCFW}}_{8,{\rm NMHV}}\,\,&=\phantom{+}(1)(2)(3) + (1)(2)(5) + (1)(2)(7) + (1)(4)(5) + (1)(4)(7)\\
\,\,&\phantom{\,=}+(1)(6)(7) + (3)(4)(5) + (3)(4)(7) + (3)(6)(7) + (5)(6)(7).
\end{split}
\end{equation}
We remind the reader of a fact that will be important repeatedly: residues are naturally alternating in the arguments, so that e.g. $(i_1)(i_2)= - (i_2)(i_1)$.
The P(BCFW) form of the amplitudes has exactly the same form as BCFW, but switching the role of even and odd integers:
\be
\label{nmhvcont}
M^{{\rm P(BCFW)}}_{n,{\rm NMHV}} =(-1)^{n-5}\Large\sum\normalsize\underbrace{(e_1)(o_2)(e_3)\,\,\cdots\phantom{+}}_{\text{\normalsize{$n-5$ terms}}}.
\ee

As shown in \cite{ArkaniHamed:2009dn}, the equality $M^{{\rm BCFW}} = M^{{\rm P(BCFW)}}$ is a (quite non-trivial) consequence of global residue theorems, which further guarantees the cyclic invariance of the amplitude.

This presentation of the NMHV amplitudes makes all of its symmetries manifest, and is strikingly ``combinatorial" in nature. One thing that is seemingly {\it not} manifest, however, is that this object has anything whatsoever to do with a local space-time Lagrangian! Each term individually has ``non-local" poles, which magically cancel in the odd/even/odd combination defining the amplitude. The cancelation of these non-local poles can be understood indirectly by the equality $M^{{\rm BCFW}} = M^{{\rm P(BCFW)}}$, since the non-local poles appearing in the two forms turn out to be different. However, this is very far from establishing that this object comes from a local Lagrangian, and one would certainly like to see the emergence of space-time in a much more direct and explicit way.

In this note, we will argue that the local space-time description of tree scattering amplitudes is actually hiding in plain sight in the BCFW sum over residues in the Grassmannian. We will show that a very natural and canonical contour deformation converts the BCFW form of tree scattering amplitudes to the CSW/Risager expansion, which is a direct reflection of the space-time Lagrangian in light-cone gauge!
\section{Brief Review of CSW and Risager}

To set the stage, let us quickly review the story of the CSW recursion relations \cite{Cachazo:2004by,Cachazo:2004kj,Cachazo:2004zb} and the very closely-related Risager recursion relations \cite{Risager:2005vk,Risager:2008yz}. 
The CSW rules are simply Feynman rules \cite{Mansfield:2005yd}, except that the vertices are off-shell continuations of MHV amplitudes, where the $\lambda$'s for internal lines with momentum $P$ are defined by\vspace{-0.3cm}
\be
\lambda_P = P| \zeta],\vspace{-0.3cm}
\ee
where $\zeta$ is an auxiliary spinor. Note that we use a different notation for this auxiliary spinor than the usual one in the literature, $\tilde \eta$, in order to not confuse this object with the SUSY Grassmann parameters. The similarity with usual Feynman rules and the hidden Lorentz invariance of this expansion is not a coincidence: the CSW rules can be derived from the Yang-Mills Lagrangian by going to a more sophisticated version of light-cone gauge \cite{Mansfield:2005yd,Gorsky:2005sf}; 
the auxiliary spinor $ \zeta$ is associated with the light-like direction defining the light-cone gauge. As usual in light-cone gauge, we have only physical degrees of freedom, the two polarizations $\pm$ of the gluons. There are cubic interactions $(+ + -)$, $(--+)$ and the quartic interaction $(++--)$. From this, it is possible to make a field redefinition to remove the anti-MHV $(++-)$ interaction; this forces the introduction of an infinite number of new MHV vertices, which must---on-shell---reproduce the MHV amplitudes. The resulting Lagrangian is precisely the one that gives the CSW rules. The equivalence between the MHV rules in a light-cone gauge and usual Lorentz-invariant formulation of the (super) Yang-Mills Lagrangian ${\cal L} = -\frac{1}{4} {\rm tr} F_{\mu \nu}^2 + \cdots$ was nicely established in a different way in \cite{Boels:2007qn}. Beginning with a twistor
space action with a large amount of gauge symmetry, one gauge-fixing leads to the usual manifestly Lorentz-invariant Yang-Mills action, while a different gauge-fixing yields the MHV Lagrangian in light-cone gauge. Thus, the CSW rules should be thought of as directly reflecting the Yang-Mills Lagrangian in light-cone gauge, encoding local space-time physics in the most succinct possible way.

For future reference, we remind the reader that the terms in the CSW expansion of the N$^{k\smallminus2}$MHV amplitude are localized on $(k-1)$ intersecting lines in the $\mathcal{Z}$-twistor space: the MHV vertices in the CSW diagrams are associated with lines in twistor space, while the internal lines are associated with points where these lines intersect. Thus, a general term in the CSW expansion of NMHV amplitudes with particles $m,k,$ and $l$ of negative helicity is localized in twistor space as shown below.
\begin{figure}[h]
\centering\includegraphics[scale=1]{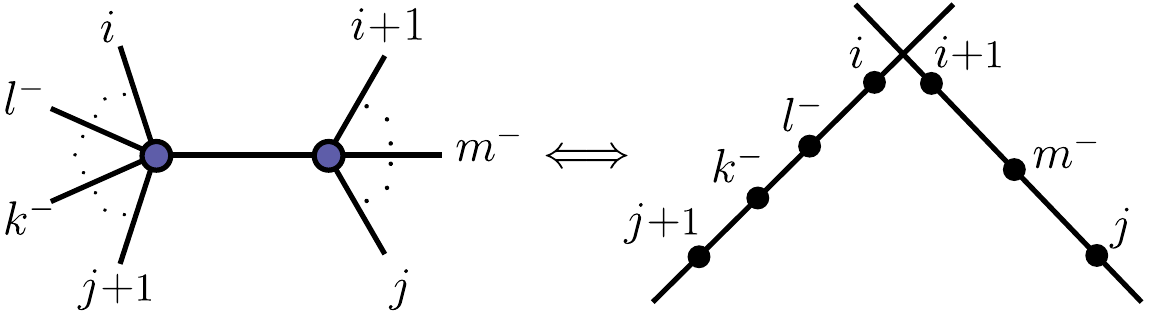}
\end{figure}

The Risager deformation is closely related, providing an alternate derivation of the CSW rules that closely parallels the logic leading the BCFW recursion relations \cite{Britto:2004ap,Britto:2005fq,Cheung:2008dn,Mason:2009sa}. As with BCFW, it involves a deformation of the spinor helicity variables; specifically, it begins by canonically deforming the $\tilde \lambda_i$'s for all the negative helicity particles:
\be\label{risager_deformation}
\tilde \lambda_i \to \tilde \lambda_i + \alpha_i z \zeta.
\ee
In order to conserve overall momentum, the $\alpha_i$ must satisfy the constraint
\be
\sum_i \alpha_i \lambda_i = 0.
\ee
Thus, for $k$ negative helicity gluons, the most general Risager deformation is labeled by $(k-2)$ parameters. It is possible to show that under this deformation the amplitude vanishes as $z \to \infty$, so that the familiar BCFW logic leads to recursion relations (see, e.g. \cite{ArkaniHamed:2008yf,Cheung:2008dn}). Remarkably, Risager showed that repeated recursive use of this deformation leads to the CSW rules \cite{Risager:2005vk}.

Below we will study the Risager expansion for $\overline{{\rm MHV}}$ amplitudes in the split-helicity configuration. In this case, the Risager diagrams consist only of ones with a three-point vertex and the lower-point $\overline{{\rm MHV}}$ amplitude connected by a propagator. We will find it useful to look at Risager deformations in momentum-twistor variables $\mu_a$, for which the general N$^{k\smallminus2}$MHV split helicity amplitude \mbox{$A(1^-,2^-,\ldots,(k\smallminus 1)^-,k^+,\ldots,(n\smallminus1)^+,n^-)$} takes the remarkably simple form:
\begin{equation}
\hat{\mu}_a=\left\{\begin{array}{llr}\mu_a + z\beta_a\et\hspace{0.5cm}&\mathrm{for~}a=1,\ldots,k\smallminus2&(\beta_a\,\mathrm{arbitrary})\\
\mu_a&\mathrm{for~}a=k\smallminus1,\ldots, n
\end{array}\right..\end{equation}
Note that this deforms $(k-2)$ terms, which is exactly the number of independent $\alpha$'s in (\ref{risager_deformation}). There are no constraints on the $\beta_a$ since---by construction---any choice of $\mu_a$ is guaranteed to produce $\tilde \lambda_a$'s that satisfy momentum conservation. This choice of $\beta_a$ determines the deformation of the negative helicity particles $\alpha_i$ as
\begin{equation}
\alpha_i = \frac{\langle i\,\,i\smallminus1\rangle \beta_{i\,\,\smallplus1} + \langle
i\smallplus1\,\, i\rangle \beta_{i\smallminus1}+\langle i\smallminus1\,\, i\smallplus1 \rangle
\beta_i}{\langle i\smallplus1\,\, i\rangle\langle i\,\, i\smallminus1\rangle}.
\end{equation}

\newpage
\section{Relaxing $\delta$-functions}
We now describe the contour deformation that will lead us from the BCFW contour in the Grassmannian to the space-time Lagrangian in light-cone gauge, passing through the CSW and Risager expansions of tree amplitudes. We begin with the form of ${\cal L}_{n,k}$ in momentum space. It is most convenient to use the momentum-twistor form, since this explicitly exhibits the (super) momentum-conserving $\delta$-functions in the pre-factor, and we can study instead the object ${\cal R}_{n,k}$.

There is something seemingly unnatural in the expression for ${\cal R}_{n,k}$: it is a nice, holomorphic contour integral, but it has explicit $\delta$-function factors! This is not unnatural at all, since these are in fact to be thought of ``holomorphic" $\delta$-functions, which are properly interpreted as poles. In other words, we may interpret $\delta^2(\mu)$ as being really
\be
\delta^2(\mu) = \frac{1}{\mu_1}\times \frac{1}{\mu_2};
\ee
or more generally, introducing a pair of auxiliary spinors $\chi,\et$, we write
\be
\label{delta}
\delta^2(\mu) = \frac{[\chi \,\et]}{[\chi\, \mu] [\et\, \mu]}
\ee
where we also demand that the contour of integration enforce the poles where \mbox{$[\chi\, \mu] = [\et\, \mu] = 0$}. Note that the expression in \mbox{equation (\ref{delta})} is not manifestly Lorentz invariant---but of course the residue obtained on the pole of both factors {\it is} Lorentz invariant. The reason for using the notation $``\delta^2(\mu)"$ is to emphasize the Lorentz invariance of the final object. Thus, when we say that the expression for ${\cal R}_{n,k}$ is a contour integral in $(k-2)(n-k-2)$ variables, we really mean that we started with a larger $(k-2)(n-k+2)$-dimensional integral and have already fixed part of the contour by specifying that it enforces $4(k-2)$ poles associated with the Bosonic parts of the $\delta^4(D_{\hat\alpha a} \mathcal{Z}_a)$-factors. Similarly, what we have been referring to as ``the" residues of ${\cal R}_{n,k}$ are really particular residues in this higher-dimensional integral, evaluated on $4(k-2)$ extra poles, with an extra $(k-2)(n-k-2)$ conditions involving the minors needed to fully-specify the residue.

This way of thinking about the $\delta$-functions explicitly as poles naturally suggests something very remarkable. We can ``relax" any one of the $\delta$-functions, using a residue theorem to move the contour off one of its associated poles, and thereby express a manifestly Lorentz-invariant residue as a sum over non-Lorentz invariant terms which involve putting an extra minor to zero. Inspired by this, we will take one of the $\delta^2$-factors and replace it by
\be
\delta^2(\mu) = \delta([\et\, \mu]) \times \frac{[\chi\, \et]}{[\chi\, \mu]},
\ee
where we mean that the pole at $[\et\, \mu] = 0$ is still being enforced while we allow ourselves the freedom to deform the contour off the pole at $[\chi\, \mu] = 0$. Note that while this expression is not Lorentz-invariant away from both poles, it {\it is} independent of the choice of $\chi$. The reason is that on the zero of $[\et\, \mu] = 0$, $\mu$ is proportional to $\et$ and we may write $\mu = d\times\et$, and so $[\chi\, \et]/[\chi\, \mu] = 1/d$ is $\chi$-independent. Thus, relaxing the $\delta$-function in this way expresses a Lorentz-invariant reside as a sum over non-Lorentz invariant terms which are a function of only a single auxiliary spinor $\zeta$. Concretely, we can do this for one of the $\delta^2(D_{\hat\alpha a} \mu_a)$ factors---e.g. that of $\hat\alpha = 1$---by making the replacement
\be
\delta^2(D_{1a} \mu_a) \to \delta(D_{1a} [\et\, \mu_a]) \times \frac{[\chi\, \et]}{D_{1a}[\chi\, \mu_a]}\;
\ee
and deforming the contour off the $D_{1a} [\chi \,\mu_a]$ pole.

Clearly, this operation can be extended to relax even more $\delta$-functions; but we will see that relaxing just one $\delta$-function ``blows up" Lorentz-invariant residues into a sum of non-Lorentz invariant terms with a beautiful physical interpretation. For the NMHV case, we will see that some of the terms in the sum are {\it precisely} the ones that appear in the CSW expansion of NMHV amplitudes. This is strongly suggested---even without a direct computation---by the localization properties of these terms both in the Grassmannian and twistor space, and the precise equality can be easily verified. Other terms in the sum do not have the appropriate localization properties and are not associated with CSW terms. The CSW terms have a local space-time interpretation and are therefore free of non-local poles, while the others do contain non-local poles. In a sense our $\delta$-relaxing contour deformation has performed a particularly powerful partial fraction expansion of the residue into a sum over local and non-local pieces. Remarkably, in the sum over residues with the alternating odd/even structure of \mbox{equation (\ref{nmhvcont})}, all the non-CSW terms appear precisely twice with opposite signs and cancel in pairs, while the remaining terms are exactly the terms of the CSW expansion of the amplitude!

For $k>3$, it is easy to see that relaxing a single $\delta$-function can not directly produce CSW terms. Nonetheless, such a canonical operation must have a physical meaning, and the only natural candidate for a non-manifestly Lorentz invariant form of amplitudes depending on a single auxiliary spinor is the Risager expansion. This raises a puzzle, however, since the Risager expansion is not unique, but is labeled by $(k-2)$ degrees of freedom. We establish the precise equivalence and understand the origin of these degrees of freedom for the case of split-helicity $\overline{{\rm MHV}}$ amplitudes, where the $(k-2)$ free parameters of the Risager deformation are seen to be quite non-trivially determined by the degrees of freedom associated with the GL$(k-2)$ ``gauge symmetry" of the momentum-twistor formula.

As was shown by Risager \cite{Risager:2005vk}, a recursive application of the Risager recursion eventually yields the CSW expansion for general amplitudes. 
Although we won't pursue this direction further in this note, this strongly suggests that the CSW expansion for general amplitudes can be directly obtained by recursively relaxing many $\delta$-function factors.

\section{NMHV and CSW from $\delta$-Relaxation}
\vspace{-0.3cm}
\subsection{Preliminaries}

Let us work in the momentum-twistor picture, where\vspace{-0.2cm}
\be
{\cal L}_{n,3} = M_{MHV} \times \int \frac{d^{n-5}D_{1a}}{(1)(2)\cdots(n)} \delta^{4|4}(D_{1a} {\cal Z}_a).\vspace{-0.2cm}
\ee
Here the $1 \times 1$ minors $(j)$ are of course just single variables $D_{1j}$; we remind the reader that the linear transformation from the $G(k,n)$ to the $G(k\smallminus2,n)$ picture makes the $(k-2) \times (k-2)$ minor $(2\,3 \cdots {k\smallminus1})_{D}$ proportional to the $k\times k$ minor $(1\,2 \cdots k)_C$, so that e.g. the minor $(2)$ in the momentum-twistor picture is proportional to the minor $(1\,2\,3)$ in the $G(3,n)$ picture. For convenience we will denote the elements of the $1\times n$ matrix $D_{\hat{\alpha} a}$ as\vspace{-0.25cm}
\be
(D_1,D_2,\ldots, D_n).\vspace{-0.3cm}
\ee
In other words, we remove the index $\hat{\alpha}$ when $k=3$ since it takes a single value.

A given residue is associated with setting $(n-5)$ of the minors to zero as is obvious: after gauge-fixing any one of the $D_{a}$, setting $(n-5)$ of the $D_{a}$'s to zero allows us to use the Bosonic $\delta$-function to solve for the remaining four $D$'s. We denote this residue as $\overline{(a_1)(a_2)(a_3)(a_4)(a_5)}$, which instructs us to write all minors in cyclic order starting from $(1)$, with $(a_1),\ldots,(a_5)$ left off. As an example with $n=8$, $\overline{(2)(3)(4)(6)(7)}$ denotes the residue $(1)(5)(8)$ where the minors $(1),(5),(8)$ are set to zero. We remind the reader once again that residues of functions in several complex variables are antisymmetric objects, so that the order in which the minors are presented matters, and e.g., $(1)(5)(8) = - (5)(1)(8)$.

We will be looking at explicit gluon amplitudes in what follows, so we need to integrate over the SUSY Grassmann parameters to extract these. This is a completely straightforward exercise. We set the gluons with $a\in I$ to have negative helicity, strip-off the ordinary momentum-conserving $\delta$-function, and we write ${\cal L}_{n,k} = \delta^4(\sum_a \lambda_a \tilde \lambda_a) L_{n,k}$ with\vspace{-0.2cm}
\be
L_{n,k} = \frac{1}{{\rm vol}[{\rm GL}(k\smallminus 2)]}\int \frac{d^{(k\smallminus 2) \times n} D_{
\hat{\alpha} a}}{(1\,2\,\cdots\,k\smallminus2)(2\,3\,\cdots\, k-1) \cdots (n\,1\,\cdots\,k\smallminus 3)} ({\rm det} \tilde{D})^4 \times \delta^4(D_{\hat{\alpha} a} \mathcal{Z}_a)\vspace{-0.2cm}
\ee
where $\tilde{D}$ is a $k \times k$ matrix
\begin{equation}
\label{tildeD}
\tilde{D}_{\alpha I} = \left(
                         \begin{array}{c}
                           \hspace{0.5cm}\lambda_{\underline{\alpha}I}\hspace{0.5cm}\vspace{0.1cm} \\
                            \hline \vspace{-0.4cm} \\ \hspace{0.5cm}\hat{D}_{\alpha I} \hspace{0.5cm} \\
                         \end{array}
                       \right) \, \, {\rm with} \, \, \hat{D}_{\alpha I} = \Theta(I-1)\sum_{a=I+1}^nD_{\alpha a} \langle I~a \rangle.
\end{equation}
Here, $\Theta(x)$ is $1$ for $x> 0$ and $0$ otherwise.

Note that while in this expression particle ``1" appears to play a special role, it could be replaced by any other starting point, with all the expressions for $L_{n,k}$ agreeing on the support of the $\delta$-functions.

Returning to the $k=3$ case, a general residue is explicitly given by\vspace{-0.1cm}
\be
\overline{(a_1)(a_2)(a_3)(a_4)(a_5)} = \int \frac{dD_{a_1} \cdots dD_{a_5}}{D_{a_1} \cdots D_{a_5}} ({\rm det} \tilde D)^4  \delta^4(D_{a_1} \mathcal{Z}_{a_1} + \cdots + D_{a_5} \mathcal{Z}_{a_5});\vspace{-0.3cm}
\ee
we can relax the $\delta$-function for the $\mu$-term by making the replacement\vspace{-0.1cm}
\begin{equation}
\delta^2(D_{a}\mu_a) \to \delta(D_{
a}[\mu_a \, \et]) \times \frac{[ \chi\, \et]}{
(D_{ a}[\mu_a \, \chi])} \equiv \frac{1}{d} \delta(D_{a}[\mu_a \, \et]).\vspace{-0.3cm}
\end{equation}
Then, we can use a residue theorem to deform the contour off $D_{a} [\chi \, \mu_a] = 0$, or equivalently off $d=0$, and write\vspace{-0.1cm}
\be
\overline{(a_1)(a_2)(a_3)(a_4)(a_5)} = \sum_{\sigma\in {\mathbb{Z}_5}} \left[\overline{(a_{\sigma(1)})(a_{\sigma(2)})(a_{\sigma(3)})(a_{\sigma(4)})d}\,(a_{\sigma(5)})\right],\vspace{-0.3cm}
\vspace{-0.1cm}\ee
where the sum is over cyclic permutations of $\{1,2,3,4,5\}$. For example, $\left[\overline{(a_1)(a_2)(a_3)(a_4)d}\,(a_5)\right]$ is given by
\begin{equation}
\int\limits_{D_{a_5} = 0} \frac{dD_{a_1} \cdots dD_{a_5}}{D_{a_1} \cdots D_{a_5}} ({\rm det} \tilde D)^4  \frac{1}{d} \delta^2(D_{a_1} \lambda_{a_1} + \cdots + D_{a_5} \lambda_{a_5}) \delta(D_{a_1} [ \et \, \mu_{a_1}] + \cdots
+ D_{a_5} [ \et\, \mu_{a_5}]).\vspace{-0.3cm}
\end{equation}

\subsection{Localization Properties of the Grassmannian}

Before we demonstrate the complete equivalence of the CSW expansion and the terms generated by ``blowing-up'' each residue of the NMHV contour, it is worthwhile to give an intuitive understanding of why this should work.

One of the strongest hints that there should be a direct connection between the CSW expansion and $\mathcal{L}_{n,k}$ is how the localization {\it in twistor-space} implied by CSW is mirrored by a {\it localization within the Grassmannian itself}. We can see this directly by Fourier-transforming the kinematical $\delta$-function $\delta^{4|4}(C_{\alpha a} \mathcal{W}_a)$ from the $\mathcal{W}$-twistor variables to their (ordinary) dual twistor-space variables $\mathcal{Z}$:
\be\label{Ztwistor_localization}
\prod_{\alpha=1}^k \delta^{4|4}(C_{\alpha a} \mathcal{W}_a) \to \int d^{4|4} z^{\alpha} \prod_{\alpha=1}^k \delta^{4|4}(\mathcal{Z}_a - C_{\alpha a} z^\alpha).
\ee
(These twistors $\mathcal{Z}_a$ are ordinary twistors, which are the duals of $\mathcal{W}_a$, and should not be confused with momentum-twistors.)

If we think of each column of $G(k,n)$ as projectively defining a point in $\mathbb{CP}^{k-1}$, then the vanishing of a minor of $G(k,n)$---consecutive or otherwise---is equivalent to some localization condition among these points in $\mathbb{CP}^{k-1}$. The first nontrivial example of this can be easily seen for $G(3,n)$, where a minor $(i\,\,j\,\,k)=0$ if and only if the corresponding points $i,j,$ and $k$ are collinear in $\mathbb{CP}^2$. It is not hard to see that the twistor-space ``collinearity operator'' $\epsilon_{IJKL}Z^I_iZ^J_jZ^K_k$, which vanishes whenever the  (Bosonic parts of the) twistors $Z_i$, $Z_j$, and $Z_k$ are collinear \cite{Witten:2003nn}, manifestly annihilates any residue of the Grassmannian supported where the minor $(i\,\,j\,\,k)$ vanishes. Similarly, for $k=4$, the ``coplanarity operator'' $\epsilon_{IJKL}Z^I_iZ^J_jZ^K_kZ^L_l$ which test whether $Z_i,\ldots,Z_l$ are coplanar, will annihilate any residue for which the minor $(i\,\,j\,\,k\,\,l)=0$. (Although beyond the scope of the present discussion, there are many reasons to suspect that localization in the Grassmannian is very natural and fundamental \cite{ABCT:2009z}.)

The simplest example to begin with is the 5-point NMHV(=$\overline{{\rm MHV}}$) amplitude. Of course, this amplitude is entirely fixed by the $\delta$-functions, and ordinarily no residue would be chosen at all.  Therefore, the contour deformation corresponding to relaxing the $\delta$-function gives rise to a sum over each of the $5$ minors\vspace{-0.3cm}
\be\label{5pointNMHV}
M_{5,{\rm NMHV}} = \left[\overline{(2)(3)(4)(5)d}\,(1)\right]+\left[\overline{(1)(3)(4)(5)d}\,(2)\right]+\ldots 
\equiv\sum_{j=1}^{5}\left[(j\,\,j\smallplus1\,\,j\smallplus2)\right].\vspace{-0.3cm}
\ee

From our discussion above, it is clear that the term in the expansion setting $(1\,\,2\,\,3) = 0$ forces the points $1,2$, 3 to be collinear in twistor space; it is trivial that NMHV amplitudes are all localized on a $\mathbb{CP}^2$ inside the $\mathbb{CP}^3$ of twistor space, so the line connecting $4,5$ intersects the line containing $1,2$, $3$ and thus, this term has the localization properties we expect of a CSW diagram. This is true for all the terms in (\ref{5pointNMHV}), and we can make an association with the terms setting the minors to zero and each of the CSW diagrams illustrated above.
\begin{figure}[t]
\centering\includegraphics[scale=1]{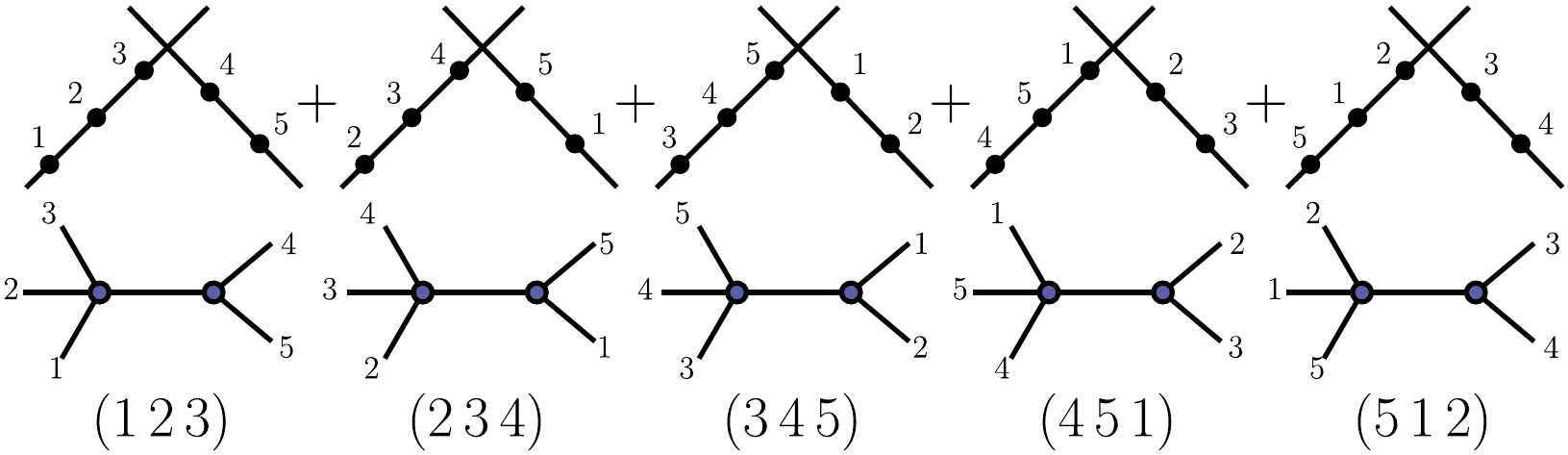}\vspace{-0.5cm}
\end{figure}

Before showing the computation that establishes the precise equivalence with the CSW terms, let us understand this localization picture for general NMHV amplitudes, starting with the 6-particle case. A given residue $(j\,\,j\smallplus1\,\,j\smallplus2)$ is blown-up into the sum of 5 terms,
\begin{equation}
(j\,\,j\smallplus1\,\,j\smallplus2) \rightarrow \sum_{k \neq j} [(j\,\,j\smallplus1\,\,j\smallplus2)(k\,\,k\smallplus1\,\,k\smallplus2)]\equiv\sum_{k\neq j}[(j)(k)]\vspace{-0.4cm}
\end{equation}
where the term $[(j)(j)]$ vanishes due to antisymmetry (or said another way, because it is a double pole with vanishing residue). Although we are choosing to write \mbox{$(j\,\,j\smallplus1\,\,j\smallplus2)\equiv(j)$} for convenience, these should not be confused with minors in the\linebreak momentum-twistor picture. Let us look at the 5 terms in the blow-up of the residue $(1\,\,2\,\,3)$; these terms have the the following localizations structure in twistor space:
\begin{figure}[h]
\centering\includegraphics[scale=1]{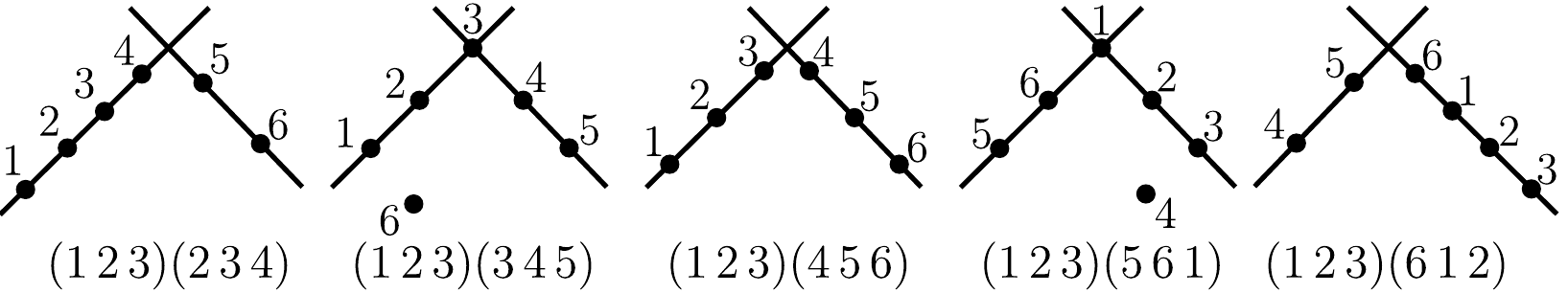}\vspace{-0.4cm}
\end{figure}

Note that while the terms $[(1)(2)],[(1)(4)],[(1)(6)]$  {\it do} have CSW localization properties, the terms $[(1)(3)]$ and $[(1)(5)]$ {\it do not}. Similarly, the terms $[(3)(1)]$ and $[(3)(5)]$ in the blow-up of $(3)$, and the terms $[(5)(1)],[(5)(3)]$ in the blow-up of $(5)$ do not have CSW localization. However, and quite remarkably, these 6 non-local terms cancel each other in pairs due to the antisymmetric property of the residues, as e.g. $[(1)(3)] + [(3)(1)] = 0$. The 9 remaining terms all have CSW localization and are indeed in perfect correspondence with the 9 CSW diagrams for this amplitude!

This pattern holds for all NMHV amplitudes. It is easiest to see this pictorially: let the sum over residues giving the BCFW form of the amplitude be represented as follows,
\be
\includegraphics[scale=1]{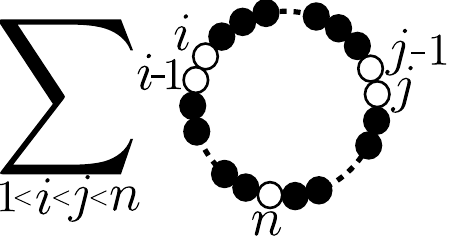}\nonumber\vspace{-0.2cm}
\ee
where each term represents \mbox{$\overline{(i\smallminus1)(i)(j\smallminus1)(j)(n)}$}, i.e., the open circles correspond to the minors that are not being set to zero.

Now, when we blow up each residue with our contour deformation, we have a sum over terms setting an extra minor tacked-on at the end of the chain to zero, which can be represented in the picture by summing over terms ``coloring-in'' one of the white dots, leaving us with 4 minors that are not set to zero. Each of these has some localization properties, but it is easy to see that the only ones that have CSW localization are the ones of the form:
\be
\includegraphics[scale=1]{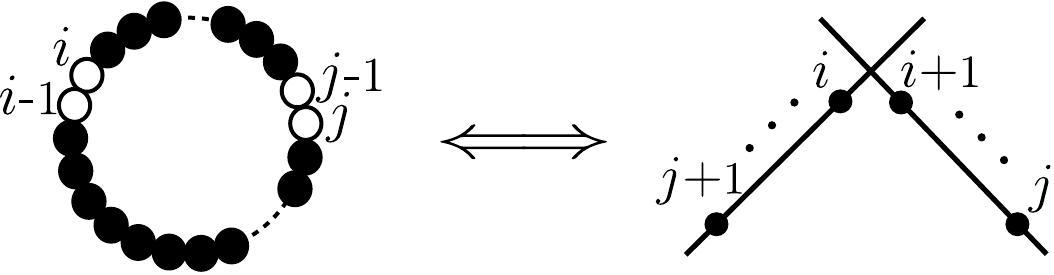} \nonumber
\ee

Now let us see what we get from coloring-in a white dot in a general term of our NMHV sum. The ones where $(n)$ is colored in automatically has good CSW properties; these give a subset of CSW diagrams, where the white circles do not include $(n)$:
\be
\includegraphics[scale=1]{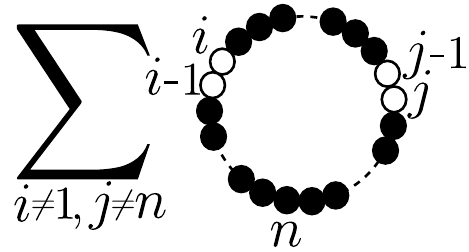} \nonumber
\ee

But in addition to these good terms, there are dangerous terms which do not have CSW localization properties, arising from coloring-in $(i-1)$; but each of these pair up with a similar term
where $(i)$ is colored in, and they cancel in pairs due to antisymmetry of the residues:
\be
\includegraphics[scale=1]{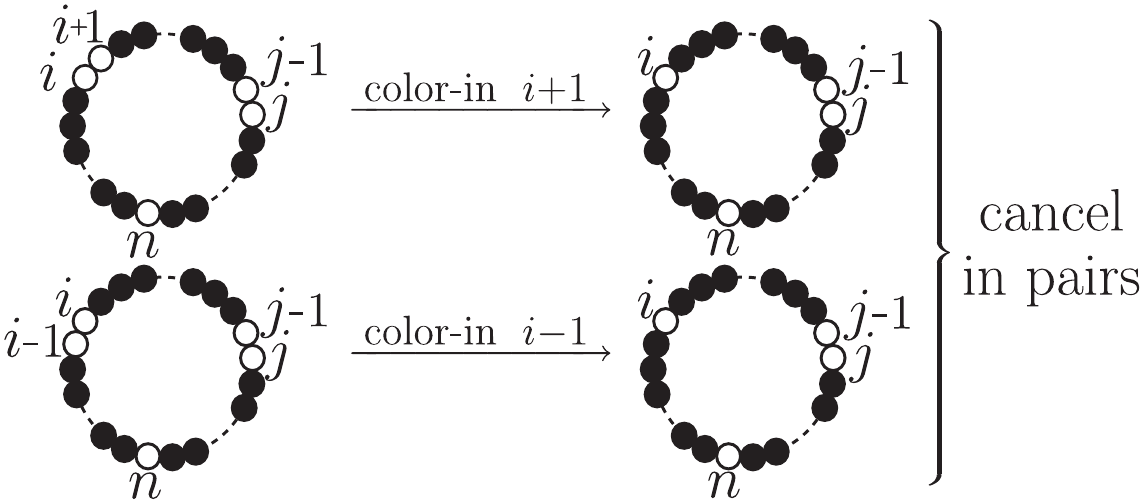} \nonumber
\ee
---with the obviously symmetrical statements holding for coloring-in $(j)$. There are also the diagrams where we color-in $(i)$ which cancel in pairs with the one where $i\to i-1$, except for the case where $i-1 = 1$, where there is no canceling diagram---but this is perfect, since the term with $i-1 =1$ (and the analogous $j=n-1$) has CSW localization
\vskip .1in
\vspace{-0.3cm}\be
\includegraphics[scale=1]{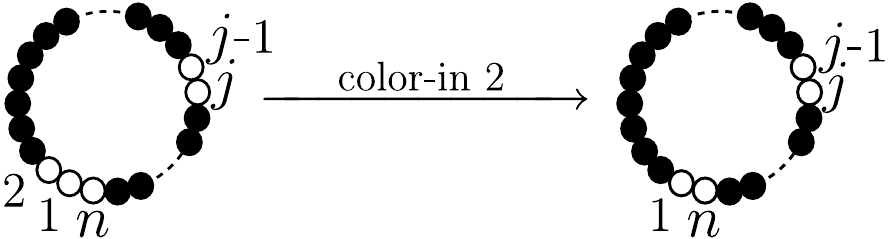} \nonumber
\vspace{-0.3cm}\ee
and provide the missing CSW terms with white circles covering $(n)$, giving us the sum over all CSW terms
\be
\includegraphics[scale=1]{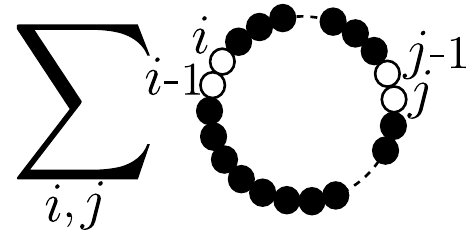} \nonumber
\ee

\subsection{Establishing the CSW Equivalence}
We finally prove that each of the remaining residues in the sum above precisely corresponds to the corresponding term in the CSW expansion of the NMHV amplitude. To begin with, it is convenient to introduce the following notation
\vspace{-0.1cm}\begin{equation}
\{ a~b~c \}  = \mu_a \langle b ~ c\rangle + \mu_b \langle c~ a\rangle + \mu_c
\langle a~b\rangle\vspace{-0.2cm}
\end{equation}
so that, e.g.,
\be
\tilde{\lambda}_i = \frac{\{i+1\,\,i\,\,i-1\}}{\langle
i+1\,\,i\rangle\langle i\,\,i-1\rangle}.
\ee

Let us compute each of the residues
$\overline{(i)(i+1)(j)(j+1)d}$, corresponding to the vanishing of all $D$'s except $D_{i}$, $D_{i+1}$,
$D_{j}$, $D_{j+1}$ and $d$.

Recall that we have three delta functions to impose:
\be
\label{sapi}
\delta^2(D_{i}\lambda_i + D_{i+1}\lambda_{i+1} + D_{j}\lambda_{j} +
D_{j+1}\lambda_{j+1}) \delta(
D_{i}[\mu_i\,\et] + D_{i+1}[\mu_{i+1}\,\et] + D_{j}[\mu_j\,\et]
+ D_{j+1}[\mu_{j+1}\,\et]).
\ee
Using GL$(1)$ to fix $D_{i}=1$, it is easy to solve explicitly for the rest of the $D$'s
\be
D_{i+1} = \frac{[\{i\,j\,j+1\}\,\et]}{[\{i+1\,\,j\,\,j+1\}\,\et]}, \; D_{j} = \frac{[\{ i\,i+1\,j+1\}\,\et]}{[\{i+1\,\,j\,\,j+1\}\,\et]}\;\, {\rm and} \;\, D_{j+1} = \frac{[\{i\,i+1\,j\}\,\et]}{[\{i+1\,\,j\,\,j+1\}\,\et]}.
\ee
Here $[\{a\, b\, c\} \, \et]$ means the Lorentz invariant contraction of spinors.

The three $\delta$-functions in (\ref{sapi}) yield a Jacobian
\be
J =
\frac{1}{[\{ i+1\,\,j\,\,j+1\} \, \et]}
\ee
while the product of $D$'s in the denominator of the residue becomes
\begin{equation}
\frac{1}{D_{i}D_{i+1}D_{j}D_{j+1}} =
\frac{[\{i+1\,\,j\,\,j+1\}\,\et]^3}{[\{i+1\,\,i\,\,j+1\}\,\et][\{i+1\,\,i\,\,j\}\,\et][\{i\,\,j\,\,j+1\}\,\et]}.
\end{equation}
Finally,
\begin{equation}
d = \langle
Z_{i}Z_{i+1}Z_jZ_{j+1}\rangle
\end{equation}
where $\langle
Z_{i}Z_{i+1}Z_jZ_{j+1}\rangle = \epsilon^{IJKL}Z_{i,I}Z_{i+1,J}Z_{j,K}Z_{j+1,L}$ is the {\it dual} conformal invariant inner product of four momentum-twistors. In fact, this particular combination has a special meaning,
\be
\frac{\langle Z_jZ_{j-1}Z_iZ_{i-1}\rangle}{\langle j\,\,j-1\rangle\langle i\,\,i-1\rangle} =(x_j-x_i)^2 = (p_i+p_{i+1}+\dots+p_{j-1})^2
\ee
which is nothing but the propagator in the corresponding CSW diagram!
\begin{figure}[h]
\centering\includegraphics[scale=1]{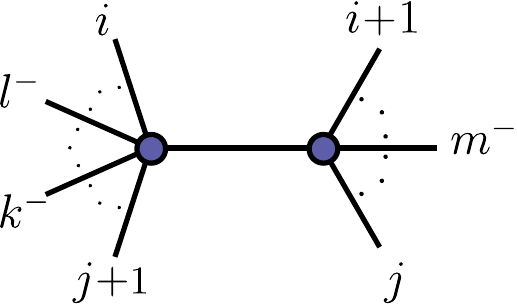}\vspace{-0.cm}
\end{figure}

In this computation we are taking as the minus-helicity particles gluons $k,l$ and $m$. Therefore, the helicity-factor $({\rm det}\,\tilde{D})$ has the form
\begin{equation}
({\rm det}\,\tilde{D}) = \left|
                         \begin{array}{ccc}
                           \lambda_m & \lambda_k & \lambda_l \\
                           \hat{D}_m & \hat{D}_k & \hat{D}_l \\
                         \end{array}
                       \right|.
\end{equation}
In the case where particle $m$ is on the right-side and $k$, $l$  on
the left-side as in the figure above, referring to equation (\ref{tildeD}), we can write
\begin{equation}
\hat{D}_m = D_{j}\langle m\,\,j\rangle + D_{j+1}\langle
m\,\,j+1\rangle = \frac{[\{j+1\,\,i\,\,i-1\}\,\et]\langle m\,\,j\rangle
- [\{j+1\,\,i\,\,i-1\}\,\et]\langle
m\,\,j\rangle}{[\{ i+1\,\,j\,\,j+1\}\,\et]}
\end{equation}
while $\hat{D}_k=\hat{D}_l=0$. Then $({\rm det}\,\tilde{D}) = \langle
k\,l\rangle\, \hat{D}_m$.

The residue $\overline{(i)(i+1)(j)(j+1)d}$, which equals $J({\rm
det}\,\tilde{D})^4/(d\, D_{i}D_{i+1}D_{j}D_{j+1})$, becomes
\be
\label{fino}
\frac{\left([\{j+1\,\,i+1\,\,i\}\et]\langle
m\,\,j\rangle - [\{j\,\,i+1\,\,i\}\et]\langle
m\,\,j+1\rangle\right)^4\langle k\,l\rangle^4}{\langle
Z_{j+1}Z_{j}Z_{i+1}Z_{i}\rangle[\{j+1\,i+1\,i\}\,\et][\{j+1\,\,j\,\,i+1\}\,\et]
[\{i\,\,j+1\,\,j\}\,\et][\{j\,\,i+1\,\,i\}\,\et]}.
\ee
A simple computation using, e.g,
\begin{eqnarray}
(p_j+\dots+p_{i+1})|i\rangle &=& \frac{\{j+1\,\,j\,\,i\}}{\langle
j+1\,\,j\rangle},\\
\langle j+1|(p_j+\dots+p_i) &=& \frac{\{j+1\,\,i\,\,i-1\}}{\langle
i\,\,i-1\rangle},
\end{eqnarray}
reveals that equation (\ref{fino}) precisely reproduces the CSW contribution associated to the corresponding diagram.

\section{Risager from $\delta$-Relaxation}
For $k>3$, it is easy to see that relaxing a single $\delta$-function does not directly lead to the CSW expansion. This is obvious since localization in the Grassmannian associated with putting $k \times k$ minors to zero for $k>3$ is not directly associated with localization on lines in twistor space. The only natural interpretation of our deformation is as the Risager expansion. An immediate question with this interpretation is precisely how the $(k-2)$ degrees of freedom of the Risager deformation are reflected in the Grassmannian picture--exactly which Risager expansion are we landing on?  In this section we establish the correspondence with Risager, and also understand the origin of the Risager degrees of freedom, by examining $\overline{{\rm MHV}}$ amplitudes. This will determine precisely which Risager expansion must be associated with our contour deformation for general $(n,k)$.

The only Risager diagrams that contribute involve the points $i,i+1$ and the internal line $P$ on one side, connected with a propagator to the lower-point $\overline{{\rm MHV}}$ amplitude on the other side
\vspace{0.1cm}\be\includegraphics[scale=1]{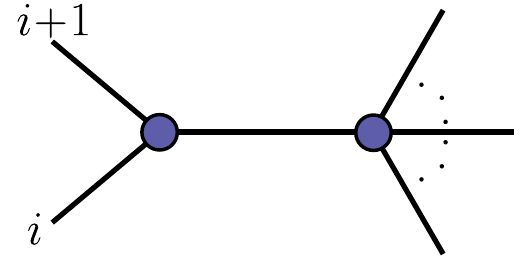} \vspace{-0.2cm}\nonumber
\ee
which can be nicely simplified to the form
\begin{equation}\label{Risterm}
A^{\rm Risager}_i = \frac{[k\,l]^4}{[\hat{1}\, \hat{2}]\dots
[\widehat{i-1}\,\,\hat{i}][i\,\,i+1][\widehat{i+1}\,\,\widehat{i+2}]\dots
[\hat{n}\,\hat{1}]}.
\end{equation}
Here, the deformation parameter $z$ is evaluated where $P^2(z^*) = 0$.
We will now see that this expansion is reproduced for the first non-trivial case of the split-helicity 6-particle $\overline{{\rm MHV}}$ amplitude
$A(1^-,2^-,3^-,4^+,5^+,6^-)$.
The $D$-matrix in the momentum twistor form of the Grassmannian is
\begin{equation}
 D = \left(
  \begin{array}{cccccccc}
    D_{11} & D_{12} & D_{13} & D_{14} & D_{15} & D_{16} \\
    D_{21} & D_{22} & D_{23} & D_{24} & D_{25} & D_{26} \\
  \end{array}
\right).
\end{equation}
As before, we will be relaxing one of the $\delta(D_{1a} \mu_a)$-factors.
Our strategy is to use four \mbox{$\delta$-function} constraints for the second row, and to solve for $D_{23},\ldots, D_{26}$ in terms of $D_{21}$ and 
$D_{22}$, and to use the remaining three $\delta$-functions to solve for
$D_{14},\ldots, D_{16}$ in terms of $D_{11}$, $D_{12}$, and $D_{13}$. Now, in deforming the contour, we will get a sum over terms where a given minor $(j)$ is set to zero. Here, we use the notation $(j)$ to refer to the minor $(j\,\,j+1\,\,\cdots\,\,j + k -3)$. We can use the condition of the vanishing of this minor to
solve for $D_{13}$ and plug it back into our equations for $D_{14},\ldots,
D_{16}$. Notice that we can gauge-fix the GL$(2)$ so that e.g. $D_{11},D_{12},D_{21},D_{22}$ are anything we like, but we will leave them arbitrary for now. The reason is that while the sum over all the terms will be GL$(2)$-invariant, each individual term will not, and as we will see the dependence on gauge degrees of freedom will precisely mirror the freedom in the Risager deformations.

A somewhat lengthy computation yields a lovely result for the term where the minor $(j)$ is set to zero; we find that it precisely corresponds to a term in the Risager expansion
\be
[(j)] = A^{\rm Risager}_{j+3}
\ee
where the Risager deformation is particularly simple and is given in terms of the following deformation on momentum twistor variables $\hat{\mu}_i = \mu_i + \beta_i z \et$ with
\be
\beta_1 = D_{22}, \quad\mathrm{and}\quad\beta_2 = D_{21}.
\ee
That is, as advertised, the degrees of the freedom in the Risager expansion are contained in the GL$(2)$ freedom of the momentum-twistor Grassmannian formula!

Moving on to the 7-point amplitude $A(1^-,2^-,3^-,4^-,5^+,6^+,7^-)$ we find exactly the same pattern: we find that the sum over terms setting a minor to zero precisely matches the Risager expansion of the amplitude, with the $\beta$-deformations now with 
\begin{equation}
\beta_1 = M_{23},\qquad \beta_2 = M_{13}, \quad\mathrm{and}\quad \beta_3 =
M_{12},
\end{equation}
where the $M_{ij}$ are determined by the GL$(3)$ gauge degrees of freedom as
\begin{equation}
M_{i,j} = \left|
            \begin{array}{cc}
              D_{2i} & D_{2j} \\
              D_{3i} & D_{3j} \\
            \end{array}
          \right|.
\end{equation}
The case for general split-helicity amplitudes follows the same pattern. We use the $D_{ij}$, $i,j=1,\dots ,n-4$, as free gauge-fixing parameters. We solve for $D_{ij}$, $i=2,\dots ,n-4$, $j=n-3,\dots ,n$ in terms of gauge-fixed
parameters $D_{ij}$, $j=1,\dots ,n-4$, and then solve for the $D_{1j}$, $j=n-2,n-1,n$ in terms of gauge
fixing parameters $D_{ij}$, $j=1,\dots ,n-4$, and $D_{1n-3}$.
Then, for each individual residue characterized by some vanishing minor
$(j)$, we determine $D_{1n-3}$, and substitute it back into
other $D_{1j}$. We can then calculate all minors and Jacobian factors, and compare with the Risager expansion.
Remarkably the two expressions agree using a Risager shift most nicely given in terms
of a deformations of $\mu$'s:
\begin{equation}
\beta_j = \left|
            \begin{array}{cccccccc}
              D_{2,1} & \dots & D_{2,j-1} & D_{2,j+1} & \dots & D_{2,n-4} \\
              \vdots & \vdots & \vdots & \vdots & \vdots & \vdots \\
              D_{n-4,1} & \dots & D_{n-4,j-1} & D_{n-4,j+1} & \dots & D_{n-4,n-4} \\
            \end{array}
          \right|.
\end{equation}
Again, the general pattern is that the deformations are constructed just
from gauge-fixing parameters. This just demonstrates the fact that
the freedom in choosing Risager deformations $\beta_j$ is included in the GL$(k-2)$ redundancy in the Grassmannian.

\section{Concluding Remarks}

We have argued that a simple and canonical ``$\delta$-relaxing" contour deformation takes us from the Grassmannian formulation of BCFW tree amplitudes---which has a remarkably ``combinatorial" form making all symmetries manifest---to the CSW expansion, which manifests the local space-time Lagrangian in light-cone gauge. Relaxing a single $\delta$-function already yields the full CSW expansion for NMHV amplitudes, and must lead to the Risager expansion for general $k$ as we established for the $\overline{{\rm MHV}}$ case. It would be interesting to see this more explicitly, and also to understand whether the recursive application of the Risager expansion leading to the CSW expansion has a natural interpretation in terms of relaxing multiple $\delta$-functions.

The operation we have found gives a natural way of ``blowing up" residues into components, separating pieces with a local space-time interpretation from the non-local ones. This allows us to give the sum over Grassmannian residues corresponding to the tree contour a ``particle interpretation" in space-time. As we will see in \cite{ABCT:2009z}, there is a second natural operation on the sum over residues---rather than blowing each residue up into many pieces, we can instead unify them together as the zero set of a single map. This manifests an even more surprising feature than a particle interpretation in space-time---the integral localizes on configurations with a ``particle interpretation" {\it in the Grassmannian}, allowing us to construct higher-point tree amplitudes by ``adding one particle at a time" to lower-point ones. Furthermore, a natural deformation not simply of the contour but of the integrand itself directly connects our Grassmannian picture with the connected prescription \cite{Roiban:2004vt} of Witten's twistor string theory \cite{Witten:2003nn,Dolan:2009wf,Berkovits:2004hg,Spradlin:2009qr}.

We find it remarkable that almost all the concepts surrounding perturbative scattering amplitudes in this decade---the twistor string theory, CSW, BCFW and Risager recursion relations, infrared equations, leading singularities and dual superconformal invariance---are unified in the Grassmannian integral we have been exploring. The only important object that has yet to make a direct appearance in this story is the light-like Wilson loop \mbox{(see e.g. \cite{Alday:2007hr,Gorsky:2009dr,Drummond:2007aua,Drummond:2007cf,Drummond:2007au,Drummond:2007bm,Drummond:2008aq})}---making this connection will surely tell us how to extract loop-level information beyond the all-loop leading singularities that are already clearly present in the Grassmannian.
\section*{Acknowledgments}

We thank Louise Dolan, Peter Goddard and Edward Witten for stimulating discussions. N.A.-H. is supported by the DOE under grant DE-FG02-91ER40654, F.C. was supported in part by the NSERC of Canada, MEDT of Ontario and by The Ambrose Monell Foundation. J.T. is supported by the U.S. Department of State through a Fulbright Science and Technology Award.

\providecommand{\href}[2]{#2}\begingroup\raggedright\endgroup

\end{document}